\documentclass[a4paper,12pt]{article}
\usepackage{graphicx,amssymb,bm,latexsym}
\newcommand{\dsty}{\displaystyle}
\newcommand{\bea}{\begin{eqnarray}}
\newcommand{\ena}{\end{eqnarray}}
\newcommand{\beq}{\begin{equation}}
\newcommand{\eeq}{\end{equation}}

\newcommand{\hs}[1]{\hspace{#1 mm}}
\renewcommand{\a}{\alpha}
\renewcommand{\b}{\beta}

\renewcommand{\d}{\delta}
\newcommand{\e}{\epsilon}

\newcommand{\pa}{\partial}
\newcommand{\nn}{\nonumber\\}
\newcommand{\p}[1]{(\ref{#1})}

\newcommand{\td}{\tilde d}

\begin{document}

\begin{titlepage}
\begin{flushright}
KU-TP 051 \\
\phantom{arXiv:yymm.nnnn}
\end{flushright}
\begin{center}
\vskip 3cm
{\Large\bf Intersecting Black Branes\vspace{2mm}\\ in Strong Gravitational Waves} \\
\vskip 10mm
{\large Li-Ming Cao$^a$, Oleg Evnin$^b$ and Nobuyoshi Ohta$^a$}
\vskip 7mm
{\em $^a$ Department of Physics, Kinki University\\
Higashi-Osaka, Osaka 577-8502, Japan}
\vskip 3mm
{\em $^b$ Institute of Theoretical Physics, Academia Sinica\\
Zh\=onggu\=anc\=un d\=ongl\`u 55, Beijing 100190, China}
\vskip 3mm
{\small\noindent  {\tt caolm@phys.kindai.ac.jp, eoe@itp.ac.cn, ohtan@phys.kindai.ac.jp}}
\end{center}
\vfill

\begin{center}
{\bf ABSTRACT}\vspace{3mm}
\end{center}

We consider intersecting black branes with strong gravitational waves propagating
along their worldvolume in the context of supergravity theories.
Both near-horizon and space-filling gravitational wave
modes are included in our ansatz. The equations of motion (originally, partial
differential equations) are shown to reduce to ordinary differential equations,
which include a Toda-like system. For special arrangements of intersecting black
branes, the Toda-like system becomes integrable, permitting a more thorough
analysis of the gravitational equations of motion.

\vfill

\end{titlepage}

\section{Introduction}

In a previous publication~\cite{CEO}, we have derived a class of gravitational
solutions corresponding to a single black brane with strong gravitational waves
propagating along its worldvolume. The algebraic structure responsible for
solving the equations of motion in that case appears to be quite general
and applies to the case of multiple intersecting branes as well. The reason
for pursuing such a generalization, which we shall presently undertake, is twofold.
First, intersecting brane solutions with (near-horizon) waves propagating on them
appear in the context of string theory studies of black hole evaporation \cite{cm}.
Second, the relative complexity of the intersecting brane case forces us to re-examine
the qualitative structure of the equation of motion and obtain a more explicit and general
picture of their properties, including a general counting of independent non-linear
gravitational wave amplitudes.

The solutions we construct exactly describe interactions of essentially non-linear structures
(black branes and strong gravitational waves). Not many solutions of this type
(black objects in time-dependent backgrounds) are known. Some analogues featuring
ordinary (rather than light-like) time dependences are black holes in
Friedmann-Robertson-Walker \cite{mcvittie,FRW} and
de-Sitter \cite{dS,Bousso:2002fq} spacetimes, as well as supergravity $p$-branes embedded
in dilaton cosmologies \cite{Maeda:2009zi}. In our present solutions, the black branes
are embedded into strong gravitational waves (rather than time-dependent cosmologies),
and a large number of arbitrary functions of the light-cone time describing the non-linear
gravitational wave amplitudes are present in the corresponding solutions.

Strong gravitational waves in flat space-time are known to be described by the metric
(see, e.g., Appendix A of \cite{Blau:2008bp})
\beq
ds^2=-2du\,dv+K_{ij}(u)x^ix^jdu^2+(dx^i)^2,
\label{brink}
\eeq
where $K_{ij}(u)$ represent the profiles of different polarization components of the wave.
In pure gravity, $K_{ij}(u)$ is constrained by $K_{ii}=0$, giving the same number
of polarizations as in linearized theory (a traceless symmetric tensor in
$D-2$ dimensions, with $D$ being the number of dimensions of space-time).
If a dilaton is present, $K_{ii}$ does not vanish and is related to the dilaton,
which gives an additional independent polarization component. The metric (\ref{brink})
is given in the so-called Brinkmann coordinates (which are not prone to coordinate
singularities and hence useful for global considerations). By a $u$-dependent
rescaling of $x^i$, it can be brought to the so-called Rosen form, in which the metric
only depends on $u$, making the planar nature of the wave front manifest.

Gravitational solutions of increasing generality describing black branes embedded in
plane waves (\ref{brink}) have been constructed in a series of publications over the last
few years. Thus, solutions for supersymmetric $p$-brane-plane-wave configurations with
specific choices of the plane wave profile have been constructed in
\cite{intersect}--\cite{MOTW1}.
A simple class of such solutions (for which the dependences on the distance from the brane
and on the light-cone time factorize in the metric) has also been considered in
\cite{lightlike} in conjunction with AdS/CFT approach to light-like `cosmologies'.
Extremal supersymmetric solutions with an arbitrary gravitational wave profile were
constructed in \cite{CDEG}, and solutions with an arbitrary profile featuring extremal
intersecting $p$-branes were constructed in \cite{MOTW2}. Non-extremal solutions for black
branes embedded in plane waves were constructed in \cite{CEO}. These latter solutions feature
strong gravitational waves localized around the black brane worldvolume, as well as
space-filling waves of the type (\ref{brink}). We shall presently generalize
the considerations of \cite{CEO} to the case of intersecting branes. (For some related
literature, see \cite{blackstring}--\cite{Narayan}.)

We have already remarked in \cite{CEO} on the motivation for investigating our present
class of solutions. Black-brane-plane-wave configurations appear in the context
of string theory investigations of black hole evaporation \cite{cm} and AdS/CFT approach
to light-like `cosmologies' \cite{lightlike}. Intersecting brane configurations
with localized near-horizon gravitational waves are particularly important in relation
to \cite{cm}. Even apart from any immediate connections to the contemporary research topics,
we believe exact non-linear solutions to be of interest in their own right, as they
give us a glimpse of the inner workings of classical higher-dimensional gravities
that is impossible to obtain by other means.

The paper consists of two main parts. First, we give a general consideration of
an assortment of intersecting black branes with gravitational waves
propagating along them. While doing so, we develop a rather systematic treatment
of the equations which displays their qualitative structure more explicitly
than the previously published investigations. Thereafter, we give a more thorough
analysis of a sub-class of such intersecting brane solutions, for which
constraints are imposed on the way the branes intersect so that the equations of
motion become solvable.

\section{General considerations}

The low-energy effective action for the supergravity system coupled
to dilaton and $n_A$-form field strengths is given by
\bea
I = \frac{1}{16 \pi G_D} \int d^D x \sqrt{\mathstrut-g} \left[
 R - \frac12 (\pa \Phi)^2 - \sum_{A=1}^m \frac{1}{2 n_A!} e^{a_A \Phi}
 F_{n_A}^2 \right],
\label{action}
\ena
where $G_D$ is the Newton constant in $D$
dimensions and $g$ is the determinant of the metric. The last term
includes both RR and NS-NS field strengths, and $a_A = \frac12
(5-n_A)$ for RR field strength and $a_A = -1$ for NS-NS 3-form. In
the eleven-dimensional supergravity, there is a four-form and no
dilaton.
We set the fermions and other background fields to zero.

\subsection{Metric and field equations}

{}From the action (\ref{action}), one can derive the field equations
\bea
R_{\mu\nu} = \frac12 \pa_\mu \Phi \pa_\nu \Phi + \sum_{A}
\frac{1}{2 n_A!} e^{a_A \Phi} \Biggl[ n_A \left( F_{n_A}^2 \right)_{\mu\nu}
 - \frac{n_A -1}{D-2} F_{n_A}^2 g_{\mu\nu} \Biggr],
\label{Einstein}
\ena
\bea
\Box \Phi = \sum_{A} \frac{a_A}{2 n_A!}
e^{a_A \Phi} F_{n_A}^2,
\label{dila}
\ena
\bea
\pa_{\mu_1} \left(\sqrt{\mathstrut- g} e^{a_A \Phi} F^{\mu_1 \cdots \mu_{n_A}} \right)
= 0 \,, \label{field}
\ena
where $F_{n_A}^2$ denotes $F_{\mu_1 \cdots \mu_{n_A}}F^{\mu_1 \cdots \mu_{n_A}}$
and $(F_{n_A}^2)_{\mu\nu}$ denotes $F_{\mu\rho\cdots\sigma}F_\nu^{~\rho\cdots\sigma}$.

The Bianchi identity for the form field is given by
\bea
\pa _{[\mu} F_{\mu_1 \cdots \mu_{n_A}]} =0.
\label{bianchi}
\ena

In this paper we  assume  the following metric form:
\bea
ds_D^2 &=& e^{2 \Xi(u,r)} \left[-2dudv + K(u,y^\a, r) du^2\right] +
\sum_{\a=1}^{d-2} e^{2 Z_\a(u,r)} (dy^\a)^2 \nn
&& +e^{2B(u,r)}\left(dr^2 + r^2 d\Omega_{\td +1}^2\right),
\label{met}
\ena
where $D=d+\tilde d+2$, the coordinates $u$, $v$ and $y^\a, (\a=1,\ldots, d-2)$
parameterize the $d$-dimensional worldvolume  where the branes
belong, and the remaining $\tilde d + 2 $ coordinates $r$ and angles
are transverse to the brane worldvolume, $d\Omega_{\tilde d+1}^2$ is
the line element of the $(\td+1)$-dimensional sphere. (Note that $u$
and $v$ are null coordinates.) The metric components $\Xi, Z_\a, B$
and the dilaton $\Phi$ are assumed to be functions of $u$ and $r$,
whereas $K$ depends on $u,y^\a$ and $r$. We shall denote derivatives with respect to $u$ and $r$ by dot and prime,
respectively. For the field strength
backgrounds, we take
\bea
F_{n_A} = E_A'(u,r) \, du \wedge dv \wedge
dy^{\a_1} \wedge \cdots \wedge dy^{\a_{q_A -1}} \wedge dr,
\label{eleb}
\ena
where $n_A = q_A +2$. With our ansatz, the
Einstein equations~\p{Einstein} reduce to
\bea
&& \sum_{\a=1}^{d-2} \ddot Z_\a + (\tilde d + 2) \ddot B +
\sum_{\a=1}^{d-2} {\dot Z_\a}^2 + (\tilde d + 2) {\dot B}^2 - 2 \dot
\Xi \left[ \sum_{\a=1}^{d-2} \dot Z_\a + (\tilde d + 2) \dot B
\right] \nn && +\; \frac12 \sum_{\a=1}^{d-2} e^{2(\Xi-Z_\a)}\pa^2_\a K
 + e^{2(\Xi- B)} \Bigg[ K \Xi'' + \frac12 K''
+ \Big(\Xi' K + \frac12 K'\Big) \Big(U'+\frac{\tilde d + 1}{r}\Big) \Bigg] \nn
&& \hs{20}= \sum_A\frac{D-q_A-3}{2(D-2)} e^{2(\Xi-B)} K
S_A (E'_A)^2 - \frac12 (\dot \Phi)^2,
\label{fe2}
\\
&& \dot \Xi' + \sum_{\a=1}^{d-2} \dot Z_\a'+(\tilde d +1)\dot B' -
\left[ \sum_{\a=1}^{d-2}\dot Z_\a +(\td +2)\dot B\right] \Xi' - \dot
B \sum_{\a=1}^{d-2} Z_\a' + \sum_{\a=1}^{d-2} \dot Z_\a Z_\a' \nn
&& \hs{20} +\frac12 \dot \Phi \Phi'=0,~~~~~~~~
\label{fe3} \\
&& \Xi'' + \Big( U' + \frac{\tilde d+1}{r} \Big)\Xi'
= \sum_A \frac{D-q_A-3}{2(D-2)} S_A (E'_A)^2,
\label{fe1} \\
&& Z_\a'' + \Big( U' + \frac{\tilde d+1}{r}\Big) Z_\a' = \sum_A
\frac{\d^{(\a)}_A}{2(D-2)} S_A (E'_A)^2,
\label{fe4}
\\
&& U''+B'' - \Big( 2\Xi' +\sum_{\a=1}^{d-2} Z_\a' - \frac{\tilde
d+1}{r}\Big) B' +2(\Xi')^2 + \sum_{\a=1}^{d-2} (Z_\a')^2 \nn
&& \hs{20}= - \frac12 (\Phi')^2 + \sum_A\frac{D-q_A-3}{2(D-2)}S_A
(E'_A)^2,~~~~~
\label{fe5}
\\
&& B'' + \Big(U'+\frac{\tilde d+1}{r}\Big) B' +\frac{U'}{r} = -
\sum_A\frac{q_A+1}{2(D-2)}S_A (E'_A)^2,
\label{fe6}
\ena
where $U$, $S_A$ and $\d_A^{(\a)}$ are defined by
\bea
\label{Udefinition}
U&\equiv& 2\Xi+ \sum_{\a=1}^{d-2} Z_\a +\tilde d B\,,
\\
S_A &\equiv& \exp\left[\e_A a_A\Phi - 2\left(2\Xi +\sum_{\a \in q_A}
Z_\a \right) \right],
\label{sa}
\ena
and
\bea
\d_A^{(\a)} = \left\{
\begin{array}{l}
D-q_A-3 \\
-(q_A+1)
\end{array}
\right. \hs{5} {\rm for} \hs{3} \left\{
\begin{array}{l}
y^\a \mbox{   belonging  to $q_A$-brane} \\
{\rm otherwise}
\end{array}
\right. , \ena respectively, and $\e_A= +1 (-1)$ is for electric
(magnetic) backgrounds. The sum of $\a$ in Eq.~(\ref{sa}) runs over
the $q_A$-brane components in the $(d-2)$-dimensional $y^\a$-space, for example
\bea
\sum_{\a \in q_A}
Z_\a=\sum_{\a_A=1}^{q_A-1}Z_{\a_A} \,.
\ena
(\ref{fe2})--(\ref{fe6}) are the $uu, ur, uv, \a\b, rr$ and $ab$ components of the
Einstein equations (\ref{Einstein}), respectively. The dilaton
equation~\p{dila} and the equations for the form field~\p{field}
and \p{bianchi} yield
\bea
&& e^{-U} r^{-(\tilde d+1)} (e^{U} r^{\tilde d+1} \Phi')'
= -\frac12 \sum_A\e_A a_A S_A (E'_A)^2,
\label{dil} \\
&& \Big( r^{\tilde d+1}e^{U} S_A E_A' \Big)'
= \Big( r^{\tilde d+1}e^{U} S_A E_A' \Big)^{\centerdot} = 0.
\label{fe8}
\ena

The configuration implicit in our ansatz is an assortment of intersecting
black $q_A$-branes, each of which is extended in a subset of $y^\alpha$-directions.
Furthermore, each brane is smeared over all the $y^\alpha$-directions except for
those already aligned with its worldvolume. This arrangement ensures translational
symmetry in $y^\alpha$.

The equations of motion we have presented possess a fairly special structure \cite{CDEG}
which enables their thorough analysis. Equations (\ref{fe1}-\ref{fe6}) and
(\ref{dil}-\ref{fe8}) are exactly identical to those for a $u$-independent problem
(static intersecting black branes). We shall show that these equations reduce to
a generalized Toda system (in special cases, such Toda-like equations can be explicitly
integrated) with one constraint on the total `energy' (i.e., the value of the Toda
Hamiltonian). In our context, which includes $u$-dependence, all the integration
constants of the equations without $u$-derivatives should be understood
as functions of $u$. These functions of $u$ correspond to the amplitudes of the
various non-linear gravitational waves present in our solutions. Substitution of
solutions to (\ref{fe1}-\ref{fe6}) and (\ref{dil}-\ref{fe8}) into (\ref{fe3}) leads
to constraints on the non-linear wave amplitudes and their $u$-derivatives. We shall prove
below that there is always only one such constraint. Finally, equation (\ref{fe2})
determines the $uu$-component of the metric, without introducing any further
constraints on the previously obtained solution of (\ref{fe3}-\ref{fe6}) and
(\ref{dil}-\ref{fe8}).

\subsection{Generalized Toda equations}

We shall now analyze explicitly equations (\ref{fe1}-\ref{fe6}) and (\ref{dil}-\ref{fe8})
and show that they reduce to a Toda-like system.

{}From \p{fe8}, we learn that
\bea
Q^{-1}S_A E_A' = c_A
\label{norm}
\ena
is a constant, where $Q^{-1}=r^{\td+1} e^U $. Equation (\ref{dil}) tells us that
\begin{equation}
\label{phiprime} Q^{-1}\Phi'=-\frac{1}{2}\sum_A \epsilon_A a_A c_A
E_{A} + f_{\Phi}\, .
\end{equation}
Similarly, we have
\begin{equation}
\label{Xiprime} Q^{-1}\Xi' = \sum_A \frac{D-q_A-3}{2(D-2)} c_A E_A+
f_{\Xi}\, ,
\end{equation}
and
\begin{equation}
\label{zprime} Q^{-1}Z_\a' = \sum_A \frac{\d^{(\a)}_A}{2(D-2)} c_A
E_A+ f_{\alpha}.
\end{equation}
where $f_{\Phi}$, $f_{\Xi}$ and $f_{\alpha}$ are functions of $u$.
{}From (\ref{fe1}), (\ref{fe4}) and (\ref{fe6}), we get
\begin{eqnarray}
\Big[Q^{-1} U' \Big]'+Q^{-1}\frac{\tilde{d}U'}{r}
\!\! &=&\!\! \sum_A\Bigg{\{}\Big[ 2(D-q_A-3) +
\Big(\sum_{\alpha=1}^{d-2}\d^{(\a)}_A\Big)
-\tilde{d}(q_A+1)\Big]\frac{c_AE'_A}{2(D-2)}\Bigg{\}} ~~~~ \nn
&=&\!\! 0 ,
\label{UU'}
\end{eqnarray}
where we have used the relation
\begin{equation}
\label{deltaandqa}
\sum_{\alpha=1}^{d-2}\d^{(\a)}_A =
\tilde{d}(q_A+1)-2(D-q_A-3) \, .
\end{equation}
So, from Eq.~(\ref{UU'}), we have
\begin{equation}
\Big(r^{2\tilde{d}+1}e^{U}U'\Big)'=0\, ,
\label{eU}
\end{equation}
and thus
\begin{equation}
\label{expU} e^{U}=h_U-\frac{f_U}{(2\tilde{d}) r^{2\tilde{d}}}\, ,
\end{equation}
where $f_U$ and $h_U$ are functions of $u$.

We now show that (\ref{norm})-(\ref{zprime}) can be re-written as a generalized Toda
system. The appearance of a Toda structure is not surprising in our context, since
a $2\times 2$ Toda-like system has already been observed for dyonic (static)
solutions~\cite{Toda}. In our case, there is an arbitrary number of different
form charges, and the number of Toda equations will equal the number of charges,
just like in \cite{Toda}.

Introducing a new variable $w$ by
\bea
\frac{d}{dw} = Q^{-1}\frac{d}{dr},
\ena
we find that \p{phiprime}--\p{zprime} can be integrated as
\bea
\Phi &=& -\frac12 \sum_A \e_A a_A c_A F_A + f_\Phi(u) w +h_\Phi(u), \nn
\label{sol1}
\Xi &=& \sum_A \frac{D-q_A-3}{2(D-2)} c_A F_A + f_\Xi(u) w+ h_\Xi(u), \\
Z_\a &=& \sum_A \frac{\d^{(\a)}_A}{2(D-2)} c_A F_A + f_\a(u)
w+h_\a(u), \nonumber
\ena
where we have defined
\bea
F_A \equiv \int dw E_A.
\label{FA}
\ena
Substituting these into \p{norm}, we obtain
\bea
\frac{d^2 F_A}{dw^2} = c_A \exp\left[ \sum_B  2c_A^{-1}
M_{AB}^{(2)}F_B + 2M^{(1)}_A c_A^{-1} w
+\left(4h_\Xi+2\sum_{\alpha\in
q_A}h_{\alpha}-\epsilon_Aa_Ah_\Phi\right)\right], \label{TodaF}
\ena
where
\bea
M_{AB}^{(2)}&=& \frac12\left\{ \frac12 \e_A a_A \e_B a_B
+2\frac{D-q_B-3}{D-2}
+\sum_{\a\in q_A}\frac{\d^{(\a)}_B}{D-2}\right\} c_A c_B, \label{M2}\\
M^{{(1)}}_{A} &=&\Bigg(2f_{\Xi}+\sum_{\alpha\in q_A}f_{\alpha}
-\frac{1}{2}\epsilon_Aa_Af_{\Phi}\Bigg)c_A.
\ena
$M_{AB}^{(2)}$ is actually a symmetric matrix, as shown in Appendix A. The terms
linear in $w$ and independent of $w$ in the exponent on the
right-hand side of (\ref{TodaF}) can be absorbed into a redefinition
of $F_A$. Indeed, introducing \beq \tilde F_A= F_A+
{M_{AB}^{(2)}}^{-1}M^{{(1)}}_{B}w+{M_{AB}^{(2)}}^{-1}c_B\left(2h_\Xi+\sum_{\alpha\in
q_B}h_{\alpha}-\frac12\epsilon_Ba_Bh_\Phi\right) \label{FAtilde}
\eeq (where summation over $B$ is understood), we obtain an
explicitly Toda-like system \bea \frac{d^2 \tilde F_A}{dw^2} = c_A
\exp\left[\sum_B 2 c_A^{-1} M_{AB}^{(2)} \tilde F_B \right].
\label{Teq} \ena These equations can be derived (as $\pa H/\pa p_A =
d\tilde F_A/dw$ and $\pa H/\pa \tilde F_A=-dp_A/dw$) from the
following Toda Hamiltonian: \bea H=\frac12 \sum_{A,B}
{M_{AB}^{(2)}}^{-1} p_A p_B -\frac12 \sum_A c_A^2 \exp\left[ \sum_B
2 c_A^{-1} M_{AB}^{(2)} \tilde F_B \right]. \label{TodHam} \ena

Generalized Toda equations have been considered extensively in mathematical
literature. They are known to be integrable \cite{Bogoyavlensky,rstn} for special
choices of $M_{AB}^{(2)}$ (constructed from roots of Lie algebras). The original Toda system
(one-dimensional non-linear `crystal' with exponential interactions between atoms)
corresponds to a sparse $M^{(2)}$-matrix ($M_{AB}^{(2)}=0$ unless $A=B\pm1$). If
$M^{(2)}$ is diagonal, the system splits into independent Liouville equations
and is particularly easily integrated. We shall analyze this case explicitly in
section 3. Any special form of the $M^{(2)}$-matrix imposes constraints on the
arrangement of the intersecting branes.

We have presently analyzed all the equations of motion without $u$-derivatives,
except for (\ref{fe5}). We shall now show that (\ref{fe5}) can be re-written
as a first order equation, which merely imposes one constraint on the integration
constants of the equations we have already considered.

Eliminating second derivatives from (\ref{fe5}) with the use of (\ref{fe6}) and (\ref{eU})
and expressing $B$ through $U$, $\Xi$ and $Z_\alpha$ according to (\ref{Udefinition}),
we get
\begin{eqnarray}
\label{reduceddoubleprime}
&&-(\tilde{d}+1)U'^2-2\tilde{d}(\tilde{d}+1)\frac{U'}{r}+2(\tilde{d}+2)(\Xi')^2
+4\Xi'\sum_{\alpha}Z'_{\alpha}+\tilde{d}\sum_{\alpha}(Z'_{\alpha})^2\nonumber\\
&&+\sum_{\alpha,\beta}Z'_{\alpha}Z'_{\beta}+\frac{\tilde{d}}{2}(\Phi')^2
=\frac{\tilde{d}}{2}\sum_A
Qc_AE'_A\, .
\end{eqnarray}
{}From the expression of $U$ given by (\ref{expU}), we find
\begin{equation}
-(\tilde{d}+1)U'^2-2\tilde{d}(\tilde{d}+1)\frac{U'}{r}
=-2\tilde{d}(\tilde{d}+1)f_Uh_UQ^{2}\,.
\label{29}
\end{equation}
Substituting (\ref{phiprime})--(\ref{zprime}) and \p{29} into
(\ref{reduceddoubleprime}), we get
\begin{equation}
\label{Eequation} \sum_{A,B}M^{(2)}_{AB}E_AE_B +2
\sum_{A}M^{{(1)}}_{A}E_A -\sum_{A}Q^{-1}c_AE_A' =-M^{(0)},
\label{constraint}
\end{equation}
where
\begin{equation}
\frac{\tilde{d}}{2}M^{(0)}=2(\tilde d +2)f_{\Xi}^2+ 4
\sum_{\alpha}f_{\Xi}f_{\alpha} + \tilde{d}\sum_{\alpha}f_{\alpha}^2
+
\sum_{\alpha,\beta}f_{\alpha}f_{\beta}+\frac{\tilde{d}}{2}f_{\Phi}^2
-2\tilde{d}(\tilde{d}+1)f_Uh_U\,.
\end{equation}
Here we have also used (\ref{deltaandqa}) and the fact that
\begin{eqnarray}
\label{sumqa}
\frac{1}{D-2}\sum_{\alpha}f_{\alpha}\Big(\delta_{A}^{(\alpha)}+q_A+1\Big)
=\sum_{\alpha\in q_A}f_{\alpha} \, .
\end{eqnarray}
Re-expressing $E_A$ through $\tilde F_A$ according to (\ref{FA}) and (\ref{FAtilde}),
and passing to differentiation with respect to $w$, we discover that (up to $w$-independent
terms) the left-hand side of (\ref{constraint}) is nothing but the `energy' of the Toda
system (\ref{Teq}), i.e., the Toda Hamiltonian (\ref{TodHam}) expressed through
$\tilde F_A$ and their $w$-derivatives. Hence, by the Toda `energy' conservation, the
left-hand side of (\ref{constraint}) cannot depend on $w$, and it simply imposes
one constraint on the integration constants of the Toda system (\ref{Teq}).

To summarize, we have reduced equations (\ref{fe1}-\ref{fe6}) and (\ref{dil}-\ref{fe8})
to a generalized Toda system. Such equations have been extensively studied and
they can be integrated in special cases (corresponding to special arrangements
of intersecting branes in our context). There is a number of arbitrary functions
of $u$ appearing in the solution (these functions of $u$ are integration constants
of the differential equations containing only derivatives with respect to $r$):
$f_U$, $h_U$, $f_\Phi$, $f_\Xi$, $f_\alpha$, $h_\Phi$, $h_\Xi$, $h_\alpha$ and $2N$
more functions\footnote{A certain number of integration constants of the Toda system may
be absorbed into a re-definition of $h_\Phi$, $h_\Xi$ and $h_\alpha$.} arising from
a general solution to the Toda system (where $N$ is the
number of form charges). There is one algebraic constraint on these functions
of $u$ due to equation (\ref{constraint}).

\subsection{The dot-prime equation}

Up to this point, we have shown that equations (\ref{fe1}-\ref{fe6}) and
(\ref{dil}-\ref{fe8}) fix all the $r$-dependences in the metric (except for the
$uu$-component), dilaton and the form fields, and leave a large number of arbitrary
functions of $u$ (the non-linear gravitational wave amplitudes), which we shall here
symbolically denote as $\eta_i(u)$
(the index $i$ runs over all the arbitrary functions of $u$ introduced in the preceding
derivations). In other words, after equations (\ref{fe1}-\ref{fe6}) and
(\ref{dil}-\ref{fe8}) have been solved, we obtain $B=B(r;\eta_i(u))$ (where the
$r$-dependence is completely determined) and similar expressions for the other functions
appearing in our ansatz for the metric, the dilaton and the form fields.

We now have to substitute the solutions to (\ref{fe1}-\ref{fe6}) and (\ref{dil}-\ref{fe8})
into the `dot-prime' equation \p{fe3}. Generically, since \p{fe3} only contains first
derivatives with respect to $u$, this should result in the following structure:
\beq
{\cal G}(r;\eta_i(u),\dot\eta_i(u))=0,
\label{Gconstr}
\eeq
where the $r$-dependence is completely fixed due to equations (\ref{fe1}-\ref{fe6}) and
(\ref{dil}-\ref{fe8}). In principle, (\ref{Gconstr}) may contain an infinite number of
constraints (one for each value of $r$) on the finite number of functions $\eta_i(u)$,
leaving no solutions. We shall now show that, due to an interplay between the structure
of \p{fe3} on the one hand and (\ref{fe1}-\ref{fe6}) and (\ref{dil}-\ref{fe8}) on the other
hand, this does not happen,
and \p{fe3} in fact always enforces only one constraint on $\eta_i(u)$ in the form
of a first-order differential equation with respect to $u$:
\beq
{\cal G}(\eta_i(u),\dot\eta_i(u))=0.
\label{G1}
\eeq
(This structure has been observed in all the previously derived solutions of this sort,
e.g., \cite{CEO,CDEG,MOTW2}, and we now infer it from the equations of motion in a general
fashion, without relying on explicit functional form of the solutions, which is how it was
seen in the previously published considerations.)
To this end, we assume that (\ref{fe1}-\ref{fe6}) and (\ref{dil}-\ref{fe8}) are satisfied
and examine the left-hand side of \p{fe3}. Differentiating the said left-hand side with
respect to $r$, and eliminating second derivatives
by use of \p{fe1}, \p{fe4} and \p{fe6}, we obtain
\bea
&& -\Big(U'+\frac{\td+1}{r}\Big)\Big[ \dot \Xi'+ \sum_{\a=1}^{d-2} \dot Z_\a'+(\td+1) \dot B'
-\Big\{\sum_{\a=1}^{d-2}\dot Z_\a+(\td +1)\dot B \Big\} \Xi' \nn
&& -\dot B \sum_{\a=1}^{d-2} Z_\a' + \sum_{\a=1}^{d-2}\dot Z_\a Z_\a'
+\frac12 \dot \Phi \Phi'\Big] +R_1 +R_2,
\ena
where
\bea
R_1 &=&
-\dot U' \Big(\Xi'+ \sum_{\a=1}^{d-2} Z_\a'\Big)
-(\td+1)\Big(\dot U' B'+\frac{\dot U'}{r}\Big)
-\Big(\sum_{\a=1}^{d-2} \dot Z_\a'+(\td+2) \dot B' \Big) \Xi'\nn
&& -\dot B' \sum_{\a=1}^{d-2} Z_\a'+\sum_{\a=1}^{d-2} \dot Z_\a' Z_\a'+\frac12\dot\Phi'\Phi'
 \nn
R_2 &=& \sum_A \frac{D-q_A-3}{2(D-2)}(S_A E_A'^2)^{\centerdot}
+ \sum_A \sum_{\a=1}^{d-2}\frac{\d_A^{(\a)}}{2(D-2)}(S_A E_A'^2)^{\centerdot}
- (\td+1) \sum_A \frac{q_A+1}{2(D-2)}(S_A E_A'^2)^{\centerdot} \nn
&&
- \Big[ \sum_{\a=1}^{d-2} \dot Z_\a+(\td+2) \dot B\Big] \sum_A\frac{D-q_A-3}{2(D-2)}S_A E_A'^2
- \dot B \sum_A \sum_{\a=1}^{d-2}\frac{\d_A^{(\a)}}{2(D-2)}S_A E_A'^2 \nn
&& + \sum_A \sum_{\a=1}^{d-2}\dot Z_A \frac{\d_A^{(\a)}}{2(D-2)}S_A E_A'^2
-\frac14 \dot\Phi\sum_A \e_A a_A S_A E_A'^2
\ena
Eliminating $B$ in favor of $U$ from $R_1$, we obtain
\bea
R_1 = \Big[-\frac{\td+1}{2\td}U'^2 -(\td+1)\frac{U'}{r}
+\frac{2}{\td}\Xi'\sum_{\a=1}^{d-2} Z_\a'+\frac{\td+2}{\td}\Xi'^2 \nn
+\frac{1}{2\td}\sum_{\a,\b}Z_\a' Z_\b' + \frac12 \sum_{\a=1}^{d-2} Z_\a'^2
+\frac14 \Phi'^2 \Big]^{\textbf{.}}.
\ena
On the other hand, with the help of \p{deltaandqa} and \p{sumqa},
$R_2$ can be reduced to
\bea
R_2 &=& -\frac12 \sum_A\Big[(S_A E_A'^2)^{\centerdot}
+S_AE_A'^2 \Big(\dot U+\frac{\dot S_A}{2S_A}\Big) \Big] \nn
&=& -\frac14 \sum_A (c_A Q E_A')\rule{0mm}{4mm}^{\centerdot} ,
\ena
where we have also used \p{fe8} and \p{norm} in deriving the second equality.
Consequently $R_1+R_2$ vanishes by \p{reduceddoubleprime}.

{}From this analysis, we conclude that, if (\ref{fe1}-\ref{fe6}) and (\ref{dil}-\ref{fe8})
are satisfied,
\bea
&& \Big[ \dot \Xi'+ \sum_{\a=1}^{d-2} \dot Z_\a'+(\td+1) \dot B'
-\Big\{\sum_{\a=1}^{d-2}\dot Z_\a+(\td +1)\dot B \Big\} \Xi'
-\dot B \sum_{\a=1}^{d-2} Z_\a' + \sum_{\a=1}^{d-2}\dot Z_\a Z_\a'
+\frac12 \dot \Phi \Phi'\Big]' \nn
&&= -\Big(U'+\frac{\td+1}{r}\Big)\Big[ \dot \Xi'+ \sum_{\a=1}^{d-2} \dot Z_\a'+(\td+1) \dot B'
-\Big\{\sum_{\a=1}^{d-2}\dot Z_\a+(\td +1)\dot B \Big\} \Xi'
-\dot B \sum_{\a=1}^{d-2} Z_\a'\nn
&& \hs{40} + \sum_{\a=1}^{d-2}\dot Z_\a Z_\a'+\frac12 \dot \Phi \Phi'\Big].
\ena
Hence, the left-hand side of (\ref{fe3}) has the form
\bea
&& \dot \Xi'+ \sum_{\a=1}^{d-2} \dot Z_\a'+(\td+1) \dot B'
-\Big\{\sum_{\a=1}^{d-2}\dot Z_\a+(\td +1)\dot B \Big\} \Xi'
-\dot B \sum_{\a=1}^{d-2} Z_\a' \nn
&& \hspace{3cm}+ \sum_{\a=1}^{d-2}\dot Z_\a Z_\a'
+\frac12 \dot \Phi \Phi' =\frac{{\cal F}(u)}{r^{\td+1} e^U}.
\label{lefthside}
\ena
Since (\ref{lefthside}) only contains first derivatives with respect to $u$, and since
$u$ only enters the metric and the dilaton through $\eta_i(u)$,
${\cal F}(u)$ can be expressed through $\eta_i(u)$ and their first derivatives, i.e.,
${\cal F}(u)={\cal G}(\eta_i(u),\dot\eta_i(u))$. Hence, (\ref{fe3}) takes the form
(\ref{G1}), as previously claimed.

\subsection{The K-equation and non-linear wave counting}

By use of (\ref{fe1}), (\ref{fe2}) becomes
\begin{eqnarray}
&&\sum_{\a=1}^{d-2}\Big[ \ddot Z_\a - 2 \dot \Xi
 \dot Z_\a+ {\dot Z_\a}^2 \Big] + (\tilde d + 2) \Big[\ddot B - 2  \dot \Xi \dot B
+  {\dot B}^2 \Big]+\frac12 (\dot \Phi)^2\nonumber\\
&& \hspace{3cm}+\; \frac12 \sum_{\a=1}^{d-2} e^{2(\Xi-Z_\a)}\pa^2_\a K +\frac12
e^{2(\Xi- B)} Q\Bigg( \frac{K'}{Q}\Bigg)'=0\, .
\label{Ksimple}
\end{eqnarray}
This equation determines $K$ (and hence the $uu$-component of the metric) after
equations (\ref{fe3}-\ref{fe6}) and (\ref{dil}-\ref{fe8}), which do not contain $K$,
have been solved.

The analysis of this equation essentially does not differ from the single
black brane case considered in \cite{CEO}. It is convenient to enforce the Brinkmann
parametrization (\ref{brink}) for the plane
wave at $r\to\infty$.
Then $\Xi$, $B$ and $Z_\alpha$ go to 0 for large $r$.
For the asymptotic large $r$ plane wave, it is not difficult to include
the $\alpha\beta$ polarizations by assuming
\begin{equation}
K(u,r,y^\alpha)=k(u,r)+K_{\alpha\beta}(u)y^\alpha y^\beta.
\label{Kkr}
\end{equation}
Then, \p{Ksimple} reduces to
\begin{eqnarray}
&&\sum_{\a=1}^{d-2}\Big[ \ddot Z_\a - 2 \dot \Xi
 \dot Z_\a+ {\dot Z_\a}^2 \Big] + (\tilde d + 2) \Big[\ddot B - 2  \dot \Xi \dot B
+  {\dot B}^2 \Big]+\frac12 (\dot \Phi)^2\nonumber\\
&& \hspace{3cm}+\; \frac12 \sum_{\a=1}^{d-2} e^{2(\Xi-Z_\a)} K_{\a\a} +\frac12
e^{2(\Xi- B)} Q\Bigg( \frac{k'}{Q}\Bigg)'=0\, .
\label{krsimple}
\end{eqnarray}
One then simply takes a solution for $\Xi$, $B$, $Z_\alpha$ and $\Phi$ (with all the arbitrary
functions of $u$ contained therein), specifies $K_{\alpha\beta}(u)$ and solves the above
equation for $k(u,r)$. The arbitrary functions of $u$ contained in our solution are
thus $K_{\alpha\beta}(u)$ and the functions of $u$ contained in the solution of
(\ref{fe3}-\ref{fe6}) and (\ref{dil}-\ref{fe8}) subject to $\Xi$, $B$ and $Z_\alpha$
going to 0 at large $r$.

If $\Xi$, $B$, $Z_\alpha$ and $\Phi$  are known, determining $k$ always merely amounts
to two integrations over $r$. We have thus shown that the original system of partial
differential equations (\ref{fe2}-\ref{fe6}) and (\ref{dil}-\ref{fe8}) with respect
to $r$ and $u$ reduces to:
\begin{itemize}
\item a generalized Toda system  (\ref{Teq}) of ordinary differential equations with
respect to $r$, which has been extensively studied and admits analytic solutions in
special cases;
\item an ordinary differential equation (\ref{G1}) with respect to $u$;
\item ordinary integrations over $r$ needed to solve (\ref{krsimple}).
\end{itemize}
This is an enormous simplification, and one is left with a clear picture of the kind of
non-linear gravitational waves propagating in our system. In the subsequent section,
we give a more explicit analysis to one of the special cases when the Toda system
can be integrated.

\section{A class of explicit solutions}

The Toda system (\ref{Teq}) can obviously be integrated when
$M^{(2)}$ is diagonal. (In that case, the Toda system splits
into individual Liouville equations, each of which can be solved.)
Imposing the vanishing condition on the off-diagonal components of $M^{(2)}$
introduces an intersection rule constraining the arrangement of the branes~\cite{Argurio:1997gt,Ohta:1997gw}.

To simplify the
discussion, let us change the notation as follows:
\begin{eqnarray}
M^{(2)}_{AA}=\xi_Ac_A^2\, ,\qquad M^{(1)}_A=\kappa_Ac_A\, ,
\label{Mdiag}
\end{eqnarray}
where
\begin{equation}
\xi_A= \frac{(q_A+1)(D-q_A-3)}{2(D-2)}+\frac{1}{4}a_A^2=1\, ,
\label{xi10}
\end{equation}
\begin{equation}
\kappa_A= 2f_{\Xi}+\sum_{\alpha\in q_A}f_{\alpha}
-\frac{1}{2}\epsilon_Aa_Af_{\Phi}\, .
\end{equation}
The second equality in (\ref{xi10}) holds for all supergravities including those
corresponding to ten-dimensional superstrings and
eleven-dimensional M-theory. With these specifications,
the Toda equations (\ref{Teq}) become
\beq
\frac{d^2 \tilde F_A}{dw^2} = c_A
e^{2 c_A \tilde F_A},
\label{Liou}
\eeq
which is solved by
\beq
\tilde F_A = -c_A^{-1}\ln\left(\frac{\sinh\left(c_A\sigma_A(w-\omega_A(u))\right)}
{\sigma_A}\right),
\label{Fsoln}
\eeq
where $\sigma_A(u)$ and $\omega_A(u)$ are integration constants of the Liouville equations
(\ref{Liou}).

{}From (\ref{FAtilde}) and (\ref{Mdiag}),
\beq
F_A=\tilde F_A-c_A^{-1}(\kappa_A w +N_A),
\label{Ftildediag}
\eeq
where we have defined
\begin{equation}
N_A=2h_{\Xi}+\sum_{\alpha\in
q_A}h_{\alpha}-\frac{1}{2}\epsilon_Aa_Ah_{\Phi}\, .
\end{equation}
Considering (\ref{constraint}), we find
\beq
\begin{array}{l}
\dsty
-M^{(0)}=\sum_A\left(c_A^2\left(\frac{dF_A}{dw}\right)^2+2c_A \kappa_A \frac{dF_A}{dw}
-c_A\frac{d^2F_A}{dw^2}\right)\vspace{2mm}\\
\dsty\hspace{3cm}=\sum_A\left(c_A^2\left(\frac{d\tilde F_A}{dw}\right)^2-\kappa_A^2
-c_A^2e^{2 c_A \tilde F_A}\right)=\sum_A\left(c_A^2\sigma_A^2-\kappa^2_A\right),
\end{array}
\label{diagconstr}
\eeq
which gives one algebraic constraint on the integration constants of the Toda system,
as it should per our general discussion in section 2. One can recognize the energy
of (\ref{Liou}) among the expressions in (\ref{diagconstr}).

We now turn to the `dot-prime' equation (\ref{fe3}).
Eliminating $B$ from (\ref{fe3}) by the use of (\ref{Udefinition}),
we obtain
\begin{eqnarray}
&&-\frac{\td + 2}{\td }\big(\dot \Xi'+\dot
U\Xi'\big)-\frac{1}{\td}\Big(\sum_{\alpha} \dot
Z_{\alpha}' +\dot U\sum_{\alpha}  Z_{\alpha}'\Big)+\frac{\td + 1}{\td }\dot U'
+\frac{1}{2}\dot{\Phi}\Phi'\nonumber\\
&&+\frac{2}{\td} \Xi' \sum_{\alpha} \dot
Z_{\alpha}+\frac{2}{\td}\dot \Xi \sum_{\alpha}
Z_{\alpha}'+2\frac{\td + 2}{\td }\dot \Xi\Xi' +
\sum_{\alpha}\dot{Z}_{\alpha}Z_{\alpha}'
+\frac{1}{\td}\sum_{\alpha}Z_{\alpha}'\sum_{\beta}\dot{Z}_{\beta}=0\,
.
\end{eqnarray}
Considering
\begin{equation}
\dot \Xi'+\dot U\Xi' = Q\big(Q^{-1}\Xi'\big)^{\centerdot}\, , \qquad
\sum_{\alpha} \dot Z_{\alpha}' +\dot U\sum_{\alpha} Z_{\alpha}'
=Q\Big(Q^{-1}\sum_{\alpha} Z_{\alpha}'\Big)^{\centerdot}\, ,
\end{equation}
and (\ref{phiprime})--(\ref{zprime}), (\ref{deltaandqa}) and (\ref{sumqa}),
we have
\begin{eqnarray}
&&-\frac{\td + 2}{\td
}\dot{f}_{\Xi}-\frac{1}{\td}\sum_{\alpha}\dot{f}_{\alpha}+\frac{\td
+ 1}{\td}Q^{-1}\dot{U}'-\frac{1}{2}\sum_A c_A\dot{E}_A \nonumber\\
&&+ \frac{1}{2}\sum_{A} \Big(2\dot{\Xi}+\sum_{\alpha\in q_A}\dot{Z}_{\alpha}
-\frac{1}{2}\epsilon_Aa_A\dot{\Phi}\Big)c_AE_A +
\frac{2}{\td}\sum_{\alpha}f_{\alpha}\dot{\Xi}+2\frac{\td + 2}{\td
}f_{\Xi}\dot{\Xi}\nonumber\\
&&+\frac{2}{\td}f_{\Xi}\sum_{\alpha}\dot{Z}_{\alpha}
+\sum_{\alpha}f_{\alpha}\dot{Z}_{\alpha}
+\frac{1}{\td}\sum_{\beta}f_{\beta}\sum_{\alpha}\dot{Z}_{\alpha} +
\frac{1}{2}f_{\Phi}\dot{\Phi}=0\, .
\end{eqnarray}
Due to (\ref{deltaandqa}), \p{sol1} and (\ref{Mdiag}), we get
\begin{equation}
2\dot{\Xi}+\sum_{\alpha\in
q_A}\dot{Z}_{\alpha}-\frac{1}{2}\epsilon_Aa_A\dot{\Phi} =\dot{N}_A+
\dot{\kappa}_Aw+\kappa_A\dot{ w} +c_A\dot{F}_A\, .
\end{equation}
Further applications of (\ref{deltaandqa}), \p{sol1} and (\ref{sumqa}) yield
\begin{eqnarray}
&&\frac{2}{\td}\sum_{\alpha}f_{\alpha}\dot{\Xi}+2\frac{\td + 2}{\td
}f_{\Xi}\dot{\Xi}+\frac{2}{\td}f_{\Xi}\sum_{\alpha}\dot{Z}_{\alpha}
+\sum_{\alpha}f_{\alpha}\dot{Z}_{\alpha}
+\frac{1}{\td}\sum_{\beta}f_{\beta}\sum_{\alpha}\dot{Z}_{\alpha} +
\frac{1}{2}f_{\Phi}\dot{\Phi}\nonumber\\
&&=\frac{2}{\td}\sum_{\alpha}f_{\alpha}\dot{h}_{\Xi}+2\frac{\td +2}{\td}
f_{\Xi}\dot{h}_{\Xi}+\frac{2}{\td}f_{\Xi}\sum_{\alpha}\dot{h}_{\alpha}
+\sum_{\alpha}f_{\alpha}\dot{h}_{\alpha}
+\frac{1}{\td}\sum_{\beta}f_{\beta}\sum_{\alpha}\dot{h}_{\alpha} +
\frac{1}{2}f_{\Phi}\dot{h}_{\Phi}\nonumber\\
&&+2(\td +1) (f_Uh_U)\dot{w}+(\td
+1)(f_Uh_U)^{\centerdot}w+\frac{1}{2}M^{(0)}\dot{w}
+\frac{1}{4}\dot M^{(0)}w+\frac{1}{2}\sum_A c_A\kappa_A
\dot{F}_A\, .\nonumber\\
\end{eqnarray}
So the dot-prime equation becomes
\begin{eqnarray}
&& M+ (\td + 1) Q^{-1}\dot{U}'+2\td (\td +1)(f_Uh_U)\dot{w}+\td(\td+1)(f_Uh_U)^{\centerdot}w
\nn
&&+ \frac{\td}{2}\sum_{A}\Bigg[-c_A\dot{E}_A+\Big(\dot{N}_A+
(\kappa_Aw)^\centerdot +c_A\dot{F}_A\Big)c_AE_A\Bigg]\nonumber\\
&&+\frac{\td}{2}M^{(0)}\dot{w}+\frac{\td}{4}\dot M^{(0)}w
+\frac{\td}{2}\sum_A c_A\kappa_A\dot{F}_A=0\, ,
\end{eqnarray}
where
\begin{eqnarray}
M \!\!\! &\equiv&\!\!\! -(\td + 2)\dot{f}_{\Xi} - \sum_{\alpha}\dot{f}_{\alpha}
+ 2\sum_{\alpha}f_{\alpha}\dot{h}_{\Xi}+2(\td+ 2)f_{\Xi}\dot{h}_{\Xi}\nonumber\\
&&+ 2 f_{\Xi}\sum_{\alpha}\dot{h}_{\alpha} + \td\sum_{\alpha}f_{\alpha}\dot{h}_{\alpha}
+ \sum_{\beta}f_{\beta}\sum_{\alpha}\dot{h}_{\alpha} +
\frac{\td}{2}f_{\Phi}\dot{h}_{\Phi}\, .
\end{eqnarray}
It is easy to deduce that
\beq
Q^{-1}\dot{U}'+2\td (f_Uh_U)\dot{w}+\td (f_Uh_U)^{\centerdot}w=0\, ,
\eeq
and we get
\beq
\label{dotprime1}
M + \frac{\td}{2}\sum_{A}c_A\Bigg{\{}-\dot{E}_A+\Big(\dot{N}_A+
(\kappa_Aw)^\centerdot +c_A\dot{F}_A\Big)E_A+ \kappa_A\dot{F}_A\Bigg{\}}
+\frac{\td}{2}M^{(0)}\dot{w} +\frac{\td}{4}\dot M^{(0)}w=0\, ,
\eeq
or, using (\ref{Ftildediag}),
\beq
\begin{array}{l}
\dsty
M+\frac{\td}{2}\sum_{A}c_A\Bigg{\{}c_A\dot{\tilde{F}}_A\left(\frac{d{\tilde{F}}_A}{dw}\right)
-\left(\frac{d{\tilde{F}}_A}{dw}\right)^{\centerdot}+c_A^{-1}\left(\dot\kappa_A
-\kappa_A(\kappa_Aw)^\centerdot-\kappa_A\dot{N}_A\right)\Bigg{\}}\vspace{2mm}\\
\dsty\hspace{7cm}+\frac{\td}{2}M^{(0)}\dot{w}+\frac{\td}{4}\dot M^{(0)}w=0\, .
\end{array}
\eeq
{}From (\ref{Fsoln}),
\beq
c_A\dot{\tilde{F}}_A\left(\frac{d{\tilde{F}}_A}{dw}\right)-\left(\frac{d{\tilde{F}}_A}{dw}
\right)^{\centerdot}=c_A\sigma_A\left(\sigma_A(w-\omega_A)\right)^{\centerdot}\,.
\eeq
Hence, we get
\beq
\begin{array}{l}
\dsty
M+\frac{\td}{2}\sum_{A}\Bigg{\{}\dot\kappa_A-\kappa_A\dot{N}_A-c_A^2\sigma_A(\sigma_A
\omega_A)^\centerdot \Bigg{\}}\vspace{2mm}\\
\dsty\hspace{1cm}+ \frac{\dot{w}\td}{2}\left(\sum_A\left(c_A^2\sigma_A^2-\kappa_A^2\right)
+M^{(0)}\right) +  \frac{w\td}{4}\left(\sum_A\left(c_A^2\sigma_A^2-\kappa_A^2\right)
+M^{(0)}\right)^\centerdot=0\, .
\end{array}
\eeq
The second line vanishes by (\ref{diagconstr}). Hence, the `dot-prime' equation (\ref{fe3})
reduces to a single ordinary differential equation
\beq
M+\frac{\td}{2}\sum_{A}\Bigg{\{}\dot\kappa_A-\kappa_A\dot{N}_A
-c_A^2 \sigma_A(\sigma_A\omega_A)^\centerdot \Bigg{\}}=0,
\eeq
as per our general discussion in section 2.

\section{Conclusions}

We have considered intersecting black supergravity branes with strong gravitational
waves propagating along their worldvolume. The problem of finding the corresponding
supergravity solutions is reduced to ordinary differential equations, including
the celebrated generalized Toda system, which can be solved explicitly in special cases.

Methodologically, our treatment presents a few substantial improvements over the
previously published material. For the static intersecting black brane case (which is a
triviality in our present context, as the gravitational waves are set to zero),
our discussion generalizes the treatment of the equations without the assumptions
made in \cite{Argurio:1997gt,Ohta:1997gw}, where $M^{(2)}$ was taken to be diagonal (a convenient representation
for this matrix is given in Appendix A),
and displays a connection to generalized Toda systems in the spirit of \cite{Toda} (see also \cite{Ohta:2003uw}).
For the non-trivial case with finite amplitude gravitational waves turned on,
we give a much more thorough and general treatment of the $ur$-component of Einstein's
equations and show, relying only on the equations of motion, but not on their explicit
solutions, that it always reduces to a single ordinary differential equation.
In the previous publications \cite{CEO,CDEG,MOTW2}, this fact was demonstrated using
the explicit form of solutions. Under our present circumstances, explicit solutions are only
available in some special cases, so a general proof of the sort presented here is
indispensable to maintain analytic control over the equations of motion. It also gives
a much clearer picture of the inner working of the formalism than the derivations
of \cite{CEO,CDEG,MOTW2}.

Physically, it would be interesting to apply our solutions to the study of AdS/CFT
realizations of light-like `cosmologies' (along the lines of \cite{lightlike}) and
black hole physics (along the lines of \cite{cm}). We defer these subjects to future investigations.

Mathematically, we have seen how very similar algebraic structures enable a thorough
analysis of the gravitational equations of motion in a sequence of successively more
elaborate examples (\cite{CEO,CDEG,MOTW2} and the present publication). It would be
interesting to understand the emergence of this structure in a precise and
coordinate-independent manner. At this point, it is clear that the presence of
a null Killing vector is essential, but it is not obvious what other assumptions
one needs to make
about the geometry to create a maximally general setting in which the simplifications
(of the sort we have thus far discovered in specialized examples) do occur.

\section*{Acknowledgement}

This work was supported in part by the Grant-in-Aid for
Scientific Research Fund of the JSPS (C) No. 20540283, No.
21$\cdot$09225 and (A) No. 22244030. The work of O.E. has been supported by
grants from the Chinese Academy
of Sciences and National Natural Science Foundation of China.

\appendix

\section{Some properties of the $M^{(2)}$-matrix}

In (\ref{M2}), we have defined $M^{(2)}$ by
\beq
M_{AB}^{(2)}= \frac12\left\{ \frac12 \e_A a_A \e_B a_B +2\frac{D-q_B-3}{D-2}
+\sum_{\a\in q_A}\frac{\d^{(\a)}_B}{D-2}\right\} c_A c_B.
\eeq
We shall now show that $M^{(2)}$ is in fact symmetric by deriving a manifestly
symmetric representation of the above formula.

First, from (\ref{sumqa}), we have
\beq
\sum_{\alpha\in q_A}\delta_B^{(\alpha)}=\frac{1}{D-2}
\sum_{\alpha}\delta_{B}^{(\alpha)}\Big(\delta_A^{(\alpha)}+q_A+1\Big)
=\sum_{\alpha}\frac{\delta_{B}^{(\alpha)}\delta_A^{(\alpha)}}{D-2}
+\sum_{\alpha}\frac{\delta_{B}^{(\alpha)}}{D-2}\big(q_A+1\big)\,.
\eeq
Considering (\ref{deltaandqa}), we find
\beq
\sum_{\alpha\in
q_A}\delta_B^{(\alpha)}=\sum_{\alpha}\frac{\delta_{B}^{(\alpha)}\delta_A^{(\alpha)}}{D-2}
+\frac{1}{D-2}\big(q_A+1\big)\Big[\td(q_B+1)-2(D-q_B-3)\Big].
\eeq
Consequently,
\beq
\begin{array}{l}
\displaystyle M_{AB}^{(2)}=\Bigg\{\frac{(D-q_A-3)(D-q_B-3)}{(D-2)^2}
+\sum_{\alpha}\frac{\delta_{B}^{(\alpha)}\delta_A^{(\alpha)}}{2(D-2)^2}\vspace{1mm}\\
\displaystyle\hspace{5cm}+\frac{\td\big(q_A+1\big)\big(q_B+1\big)}{2(D-2)^2}
+\frac{1}{4}\epsilon_Aa_A\epsilon_Ba_B\Bigg\}c_Ac_B
\end{array}
\eeq
The latter expression is manifestly symmetric, which proves our claim.

\end{document}